\documentclass[twocolumn,twoside,slac_two]{revtex4}
\usepackage{graphicx}
\usepackage{fancyhdr}
\usepackage{multirow}
\begin{document}

\title{CDF Hot Topics}

%

\author{B. Casal (on behalf of the CDF Collaboration)}
\affiliation{Instituto de F\'{\i}sica de Cantabria (CSIC-Univ. Cantabria), 
Avda. de los Castros s/n, 39005 Santander, Spain}

\begin{abstract}
We present recent CDF results based on approximately 1~fb$^{-1}$ of
$p\bar p$ collisions at $\sqrt s = 1.96$~TeV delivered at Fermilab
Tevatron. Results shown include the observation of the $B_s$
oscillation frequency, the first observation of bottom baryon
$\Sigma_b^{(*)\pm}$ states, updates on $B$ hadrons lifetimes, and
searches for rare decays in the $b\to s\mu^+\mu^-$ transition and 
in charmless two-body $B$ decays. 
\end{abstract}

\maketitle

\thispagestyle{fancy}


\section{Introduction}

The Tevatron collider at Fermilab, operating at $\sqrt s = 1.96$~TeV,
has a huge $b\bar b$ production cross section which is several orders
of magnitude larger than the production rate at $e^+e^-$ colliders
running on the $\Upsilon(4S)$ resonance. In addition, on the
$\Upsilon(4S)$ only $B^+$ and $B_d^0$ are produced, while higher mass
$b$-hadrons such as $B_s^0$, $B_c$, $b$-baryons, $B^*$, and $p$-wave
$B$ mesons are currently produced only at Tevatron. In order to
exploit the possibility to study those variety of heavy $b$-hadrons in
a busy hadronic environment, dedicated detector systems, trigger and
reconstruction are crucial.

In the following subsections we briefly describe the Tevatron collider,
the CDF II detector, and the trigger strategies used at CDF for heavy
flavor physics. Then, in the following sections we discuss the most 
recent and interesting heavy flavor results at CDF.

\subsection{The Tevatron Collider}

Tevatron is a superconducting proton-synchrotron at the final stage of the
Fermilab accelerator complex. In Run II (mid-2001--present), it accelerates
36 bunches of protons against 36 bunches of anti-protons producing one crossing 
every 396~ns at $\sqrt s = 1.96$~TeV. The luminous region of the Tevatron 
beam has an RMS of $\sim30$~cm along the beamline (z-direction) with a 
transverse beamwidth of about 30~$\mu$m.

\begin{figure}[ht]
\centering
\includegraphics[width=80mm]{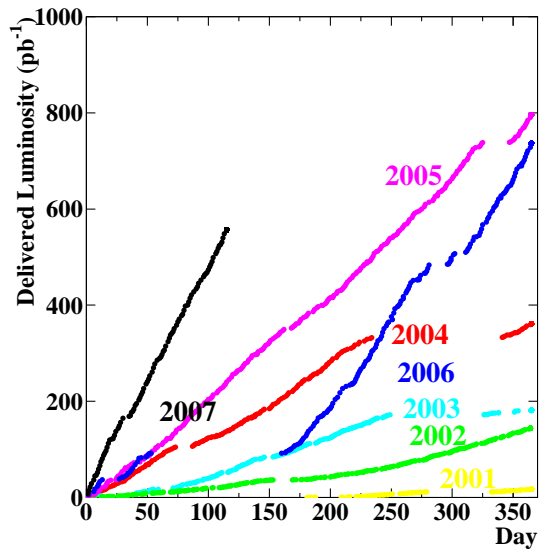}
\caption{Tevatron delivered luminosity per calendar year up to May 2007.}\label{fig:tev}
\end{figure}

The ins\-tan\-ta\-neous lu\-mi\-no\-si\-ty has been ri\-sing stea\-dily dur\-ing Run II up to the
world record peak of 2.92$\times10^{32}$~cm$^{-2}$s$^{-1}$, and regularly 
exceeds 2$\times10^{32}$~cm$^{-2}$s$^{-1}$. The machine typically delivers data 
corresponding to an integrated luminosity of $>30$~pb$^{-1}$ per week, which is
recorded with an average data-taking efficiency of about $85\%$ at CDF. The 
increase in accelerator performance throughout Run II can be seen by the
delivered luminosity per calendar year, as shown in Fig.~\ref{fig:tev}. As of
May 2007, the total integrated luminosity delivered by the Tevatron to CDF is
$\sim2.7$~fb$^{-1}$ with about $2.2$~fb$^{-1}$ recorded to tape by the CDF 
experiment. However, most results presented here are based on about 1~fb$^{-1}$
of data. Around 8~fb$^{-1}$ are expected to be delivered until the shutdown 
of the Tevatron end in 2009.

\subsection{The CDF II Detector}

The CDF II detector is a 5000 t, multipurpose, solenoidal magnetic-spectrometer
surrounded by 4$\pi$ calorimeters and muon detectors, it is axially and azimuthally
symmetric around the interaction point. Its excellent tracking performance, good 
muon coverage, and particle identification (PID) capabilities allow a broad 
flavor-physics program. We briefly outline the sub-detectors pertinent to the 
analyses described here, additional details can be found elsewhere~\cite{tdr}.

\begin{figure}[ht]
\centering
\includegraphics[width=75mm]{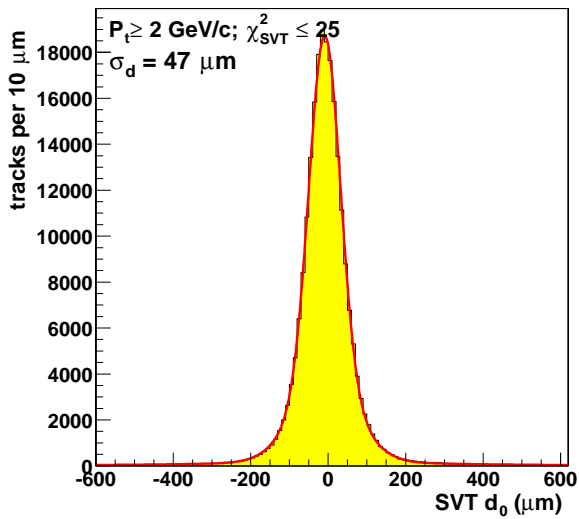}
\caption{Impact parameter resolution provided by the silicon micro-vertex detector. The
resolution is $47~\mu$m, while the typical heavy flavor trigger requires tracks with
impact parameter grater than $120~\mu$m.}\label{fig:svt}
\end{figure}

The CDF II tracking system consist of an inner silicon system surrounded
by a cylindrical gas-wire drift chamber, both immersed in a 1.4~T solenoidal 
magnetic field with 135~cm total lever arm. Six (central region, $|\eta|<1$)
to seven (forward, $1<|\eta|<2$) double-sided silicon layers, plus one 
single-sided layer, extend radially from 1.6 to 22~cm (28~cm) from the beam 
line in the central (forward) region, fully covering the luminous region.
The chamber provides 96 (48 axial and 48 stereo) samplings of charged-particle
paths between 40 and 137~cm from the beam, and within $|\eta|<1$. The long
lever arm of the tracker provides a superb mass-resolution with 
$\sigma(p_T)/p_T^2 \sim 0.1\%$ GeV$^{-1}$. In addition, silicon measurements
close to the beam allow precise reconstruction of decay vertexes, with typical
resolution of 35~$\mu$m in the transverse plane --shown in Fig. \ref{fig:svt}, 
which includes a contribution of 32~$\mu$m from the width of the $p\bar p$ 
interaction region-- and 70~$\mu$m along the beam 
direction.

Four layers of planar drift chambers detect muon candidates with 
$p_T > 1.4$~GeV/c in the $|\eta| < 0.6$ region, while conical sections of drift 
tubes extend the coverage to $0.6 < |\eta| < 1.0$ for muon candidates with 
$p_T > 2.0$~GeV/c. 

\begin{figure*}[ht]
\centering
\includegraphics[width=72mm]{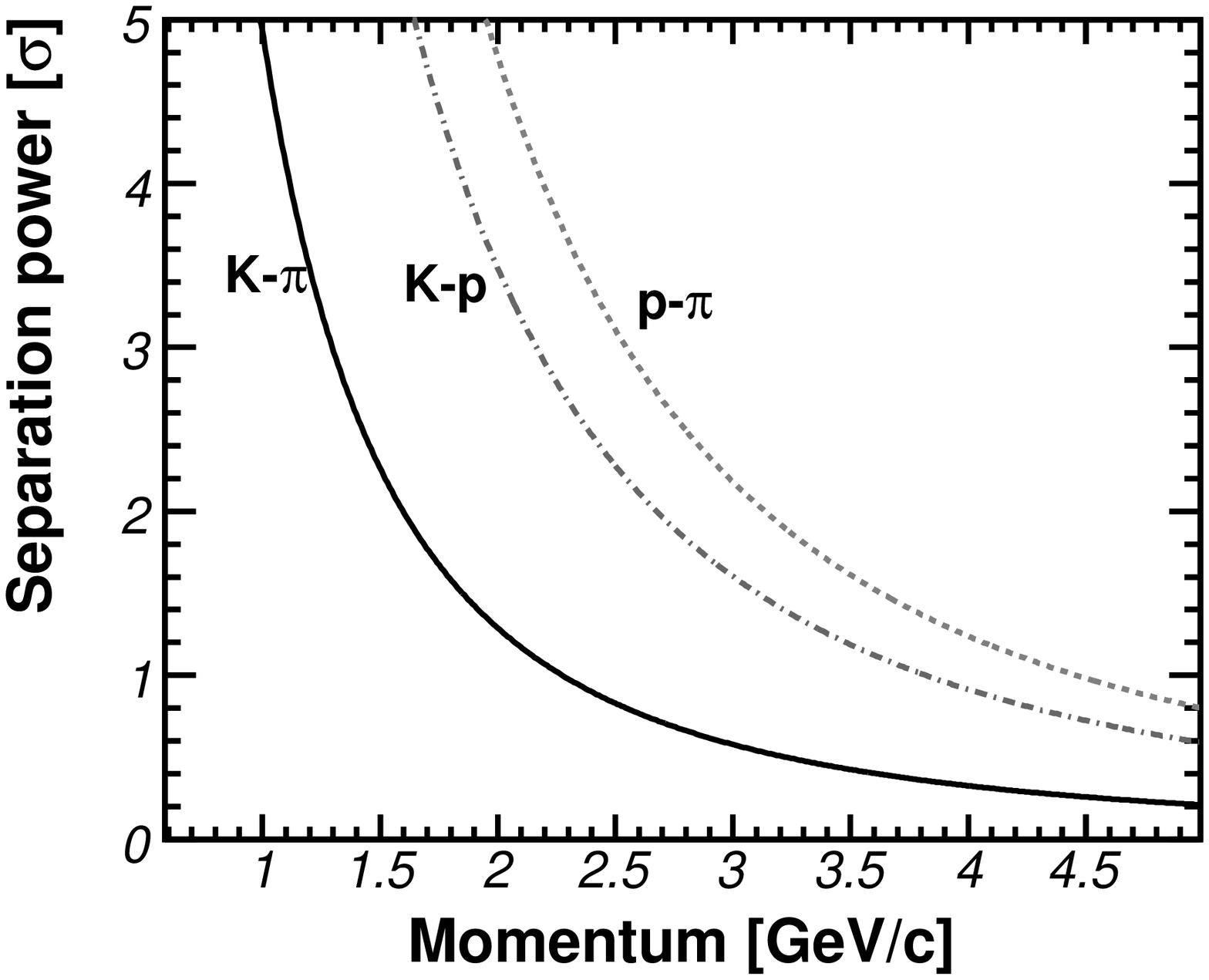}
\includegraphics[width=83mm]{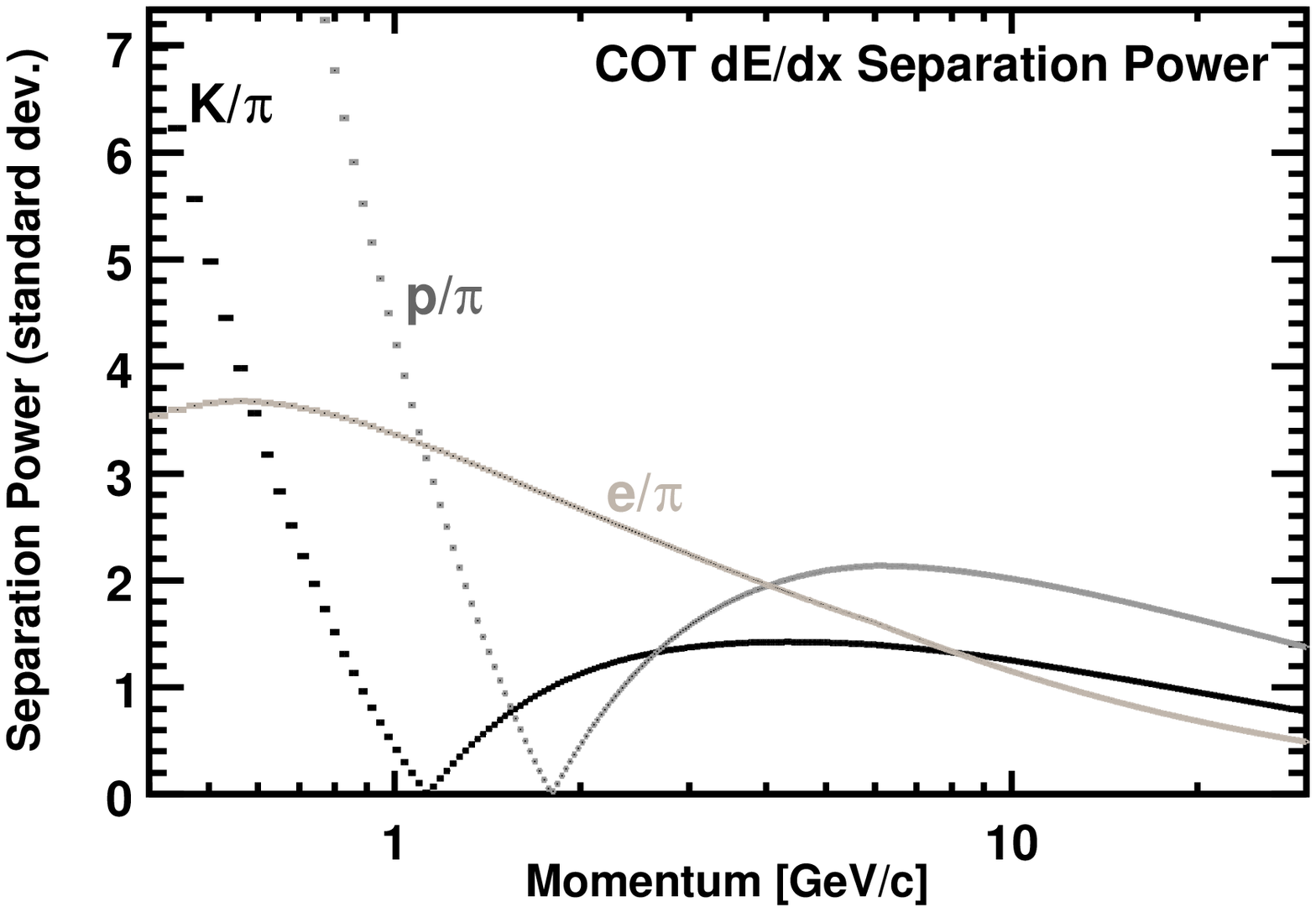}
\caption{Particle identification separation power in units of standard deviations 
provided by the TOF detector (left) and the drift chamber $dE/dx$ (right).}\label{fig:pid}
\end{figure*}

Low momentum PID is obtained with a scintillator based 
Time-of-Flight (TOF) detector with about 110~ps resolution, that provides a 
separation between kaons and pions greater than $2\sigma$ for charged particles 
with $p < 1.5$~GeV/c, see Fig. \ref{fig:pid}(left). The information of specific 
energy loss from the drift chamber ($dE/dx$) complements the PID with a nearly 
constant $1.4\sigma$  K$/\pi$  separation for higher momentum charged particles 
($pT > 2.0$~GeV/c), see Fig. \ref{fig:pid}(right).

\subsection{Trigger Strategies}

CDF exploits its unique ability to trigger events
with charged particles originated in vertexes displaced
from the primary $p\bar p$ vertex (displaced tracks) \cite{svt}. 
Displaced tracks are identified by measuring with 35~$\mu$m
intrinsic resolution their impact parameter (see Fig.~\ref{fig:svt}),
which is the minimum distance between the particle direction
and the primary $p\bar p$ vertex in the plane transverse
to the beam. Such a high accuracy can be reached
only using online the silicon information, a challenging
task that requires to read-out $212,000$ silicon channels
and to complete hit-clustering and pattern recognition
within the trigger latency. In a highly parallelized 
architecture, fast pattern matching and linearized track
fitting allow reconstruction of 2D-tracks in the plane
transverse to the beam with offline quality by 
combining drift chamber and silicon information, within
a typical latency of 25~$\mu$s per event.
Using the above device, CDF implemented a trigger 
selection that requires only two displaced tracks in the 
event, to collect pure samples of exclusive non-leptonic 
b-decays for the  first time in a hadron collider.
However, an impact-parameter based selection biases the 
decay-length distributions, and therefore a trigger efficiency 
dependence --that models the acceptance as a function of proper 
decay time-- must be taken into account for time dependent measurements.

Besides the trigger on displaced tracks, past 
experience from Run I suggests that triggering on   
final states containing single leptons or dileptons is a 
successful strategy to select samples of b-hadron decays, 
since semileptonic ($B\to\ell\nu_\ell X$) and charmonium 
($B \to J/\psi[\ell^+\ell^-]X$)  decays represent about 20\%
of b-meson widths and have relatively clean 
experimental signatures. Such a `conventional' approach 
was adapted to the upgraded detector: identification 
of muon down to low momenta allows for efficient 
dimuon triggers in which we select charmonium or 
rare decays and then we fully reconstruct several 
decay modes. On the other hand, semileptonic triggers 
require a displaced track in addition to the muon (or 
electron), providing cleaner samples. 

\section{Measurement of \boldmath{$B_s^0-\overline{B}_s^0$} Oscillation Frequency}

The precise determination of the $B_s^0-\overline{B}_s^0$ 
oscillation frequency  $\Delta m_s$ from a time-dependent 
analysis of the $B_s^0-\overline{B}_s^0$ system 
has been one of the most important goals for heavy  flavor 
physics at the Tevatron. This frequency can be used to strongly 
improve the knowledge of the Cabbibo-Kobayashi-Maskawa (CKM) 
matrix, and to constraint contributions from New Physics. 

The probability $\mathcal P$ for a $B_s$ meson produced at 
time $t = 0$ to decay as a $ B_s$ ($\overline B_s$) at 
proper time $t > 0$ is, neglecting effects from CP violation 
as well as possible lifetime difference between the heavy and 
light $B_s^0$ mass eigenstates, given by
\begin{equation}
  \mathcal{P}_\pm(t) = \frac{\Gamma_s}2e^{-\Gamma_st}\left[1\pm\cos\left(\Delta m_st\right)\right],
\label{eq:mixProb}
\end{equation}
where the subscript ``+'' (``-'') indicates that the 
meson decays as $B_s$ ($\overline B_s$). Oscillations have 
been observed and well established in the $B_d$ system. 
The mass difference  $\Delta m_d$ is measured to be 
$\Delta m_d = 0.505\pm0.005$~ps$^{-1}$ \cite{dmd}.

In the $B_s^0-\overline{B}_s^0$ system oscillation have also 
been established but till winter 2006 all attempts to measure
$\Delta m_s$ have only yielded a combined lower limit on the 
mixing frequency of  $\Delta m_s> 14.5$~ps$^{-1}$ at 95\% confidence 
level (C.L.). Indirect  fits constraint  $\Delta m_s$ to be below 
24~ps$^{-1}$ at 95\% C.L. within the standard model.
In the 2006 spring the D0 experiment presented the first double 
sided 90\% C.L. limit \cite{d0mixing} and CDF shortly afterwards 
presented the  first precision measurement on $\Delta m_s$, with 
a significance of the signal of about $3\sigma$  at that time 
\cite{cdfPRL1}. Just a few months later the CDF collaboration 
updated their result using the same data, but improved analysis 
techniques and were able to announce the observation of the 
$B_s^0-\overline{B}_s^0$ mixing frequency \cite{cdfPRL2}.

The canonical B mixing analysis proceeds as follows. The b  flavor 
($b$ or $\bar b$ of the $B$ meson at the time of decay) is determined 
from the charges of the reconstructed decay products in the final state. 
The proper time at which the decay occurred is determined from the 
transverse displacement of the $B_s$ decay vertex with respect to the 
primary vertex, and the $B_s$ transverse momentum with respect to the 
proton beam. Finally the production $b$ flavor must be known in order to 
classify the $B$ meson as being mixed (production and decay $b$ flavor 
are different) or unmixed (production and decay $b$ flavor are equal) 
at the time of its decay. 

The significance $\mathcal S$ of a mixing signal is given by:
\begin{equation} 
  \mathcal S \sim \sqrt{\frac{\epsilon D^2}{2}} \times\,
    \frac{S}{\sqrt{S+B}}\times
    \exp\left(-\frac{\Delta m_s^2\sigma_{ct}^2}{2}\right),
\label{eq:significance}
\end{equation}
where $S$ and $B$ are the signal and background event yields, respectively.
$\epsilon \mathcal D^2$ is the figure of merit for the flavor tagging, 
where $\epsilon$  is the efficiency to tag a given $B_s$ decay
candidate, and $\mathcal D = 1-P_w$ is the so-called dilution, a damping 
term which is related to 
the imperfect tagging, being $P_w$ the probability of a wrong tag.
$\sigma_{ct}$ is the proper decay time resolution, which is
crucial for this analysis especially at large $\Delta m_s$ values. 

We will in the following sections discuss those various ingredients to the 
mixing analysis --focusing in the improvements with respect to the 
analysis~\cite{cdfPRL1} presented in last year FPCP conference-- and 
then present the result.

\subsection{Signal Yields}

Several improvements with respect to the analysis in Ref~\cite{cdfPRL1}
lead to an increased $B_s$ signal yield. The decay sequences used are
the hadronic channels $\bar B_s^0\to D_s^+\pi^-,\; D_s^+\pi^-\pi^+\pi^-$ 
and the semileptonic channels $\bar B_s^0\to D_s^{+(*)}\ell^-\bar\nu_\ell$, 
$\ell =e\;\mathrm{or}\;\mu$, where $D_s^+\to\phi\pi^+$, $K^*(892)^0K^+$,
and $\pi^+\pi^-\pi^+$, and $\phi\to K^+K^-$, $K^{*0}\to K^-\pi^+$.

Particle identification techniques provided by the TOF and $dE/dx$ 
information are used to find kaons from $D_s$ meson decays, allowing 
us to relax kinematic selection requirements on the $D_s$ decay products. 
This results in increased efficiency for reconstructing the $D_s$ meson
while maintaining excellent signal to background ratio. 

In the semileptonic channel, the main gain is in the $D_s^+\ell^-$,
$D_s^+\to \bar K^*(892)^0K^+$ sequence, where the signal is increased 
by a factor of 2.2 using the particle identification techniques. 
An additional gain in signal by a factor of 1.3 with respect to our 
previous analysis comes from adding data selected with different trigger 
requirements. In total, the signal of 37,000 semileptonic $B_s$ decays 
in~\cite{cdfPRL1} is increased to 61,500, and the signal to background 
improves by a factor of two in the sequences with kaons in the final state. 

\begin{figure}[ht]
\centering
\includegraphics[width=80mm]{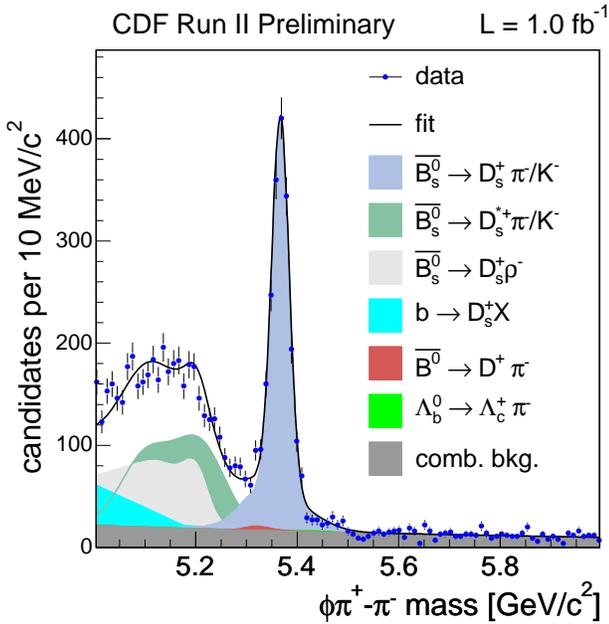}
\caption{Invariant mass distribution of  $\bar B_s^0\to D_s^{+}(\phi\pi^+)\pi^-$
candidates.}
\label{fig:golden}
\end{figure}

In the hadronic channels, we employ an artificial neural network (ANN) 
to improve candidate selection resulting in larger signal yields at 
similar or smaller background levels. The ANN selection makes it 
possible to use the additional decay sequence $\bar B_s^0\to D_s^+\pi^-\pi^+\pi^-$, 
with $D_s^+\to\pi^+\pi^-\pi^+$, as well. 
The neural network is trained using simulated signals events generated 
with Monte Carlo methods. For combinatorial background, we use sideband 
regions in the upper-mass distribution of the $B_s$ candidates from data.
We add significant statistics using the partially reconstructed hadronic 
signal between 5.0 and 5.3~GeV$/c^2$ from $\bar B_s^0\to D_s^{*+}\pi^-$,
$D_s^{*+}\to D_s^+\gamma/\pi^0$ in which a photon or a $\pi^0$ from the 
$D_s^{*+}$ is missing and $\bar B_s^0\to D_s^{+}\rho^-$, $\rho^-\to\pi^-\pi^0$
in which a $\pi^0$ is missing. The mass distribution for the highest 
statistical mode, $\bar B_s^0\to D_s^{+}(\phi\pi^+)\pi^-$, as well as for the
partially reconstructed modes is shown in Fig. \ref{fig:golden}. Table 
\ref{tab:yield} summarizes the signal yields.

With all these improvements, the statistical size of our data 
sample is increased by a factor of 2.5. 

\begin{table}[h]
\begin{center}
\caption{Signal yields ($S$) and signal to background ratio ($S/B$) in the 
hadronic decay sequences. The gain refers to the percentage increase in 
$S/\sqrt{S+B}$.}
\begin{tabular}{|l|c|c|c|}
  \hline
  Decay Sequence                           & Signal & S/B  & gain [\%]    \\ \hline
  $\bar B_s^0\to D_s^+[\phi\pi^+]\pi^-$    &  1900  & 11.3 & 13           \\\hline
  Partially reconstructed                  &  3300  & 3.4  & new          \\\hline
  $\bar B_s^0\to D_s^+[K^{*0}K^+]\pi^-$    &  1400  & 2.0  & 35           \\\hline
  $\bar B_s^0\to D_s^+[(3\pi)^+]\pi^-$     &  700   & 2.1  & 22           \\\hline
  $\bar B_s^0\to D_s^+[\phi\pi^+](3\pi)^-$ &  700   & 2.7  & 92           \\\hline
  $\bar B_s^0\to D_s^+[K^{*0}K^+](3\pi)^-$ &  600   & 1.1  & 110          \\\hline
  $\bar B_s^0\to D_s^+[(3\pi)^+](3\pi)^-$  &  200   & 2.6  & new          \\\hline
\end{tabular}
\label{tab:yield}
\end{center}
\end{table}

\subsection{Decay Length Resolution}

One of the critical input to the analysis is the proper decay 
time resolution. It is the limiting factor of the sensitivity 
of the signal at large  $\Delta m_s$ values. For setting a limit 
a too optimistic proper decay time resolution estimate could 
potentially lead to the exclusion of  $\Delta m_s$ regions we are 
actually not sensitive to. Therefore  $\sigma_{ct}$ has been 
measured directly on data. CDF exploits prompt $D$ decays plus 
tracks from the primary vertex to mimic all $B$ decay topologies 
studied in this analysis. On an event-by-event basis, the decay 
time resolution is predicted, taking into account dependences on 
several variables, such as isolation, vertex $\chi^2$, etc.
The mean $\sigma_{ct}$  for hadronic events at CDF is 26~$\mu$m and for 
semileptonic events about 45~$\mu$m.
This excellent decay length resolution is reached at CDF thanks to 
the innermost silicon layer at a distance of about 1.2~cm from the 
collision point.

\subsection{Flavor Tagging}

While the flavor of the $B_s$ candidate at decay time is unambiguously 
defined by the charges of its daughter tracks, the flavor at 
production can be inferred, with a certain degree of uncertainty, 
by flavor tagging algorithms. Two type of flavor tags can be applied: 
opposite-side and same-side flavor tags. Opposite-side tags infer 
the production flavor of the $B_s$ from the decay products of the $B$ 
hadron produced from the other $b$ quark in the event. Lepton tagging 
algorithms are based on semileptonic $b$ decays into an electron or muon
$(b\to\ell^-X)$. The charge of the lepton is thus correlated to the 
charge of the decaying $B$ hadron. Jet charge tagging algorithms use 
the fact that the charge of a $b$ jet is correlated to the charge of the 
$b$ quark. Kaon tagging are based on the CKM favored quark level decay 
sequence $(b\to b\to s)$. The charge of the kaon from opposite-side 
$B$ decays is correlated to the $b$ flavor. CDF combines these three 
tagging techniques using a Neural Network approach. The performance of 
the opposite-side flavor tagging algorithm is measured in kinematically 
similar $B_d$ and $B^+$ semileptonic samples. The $\Delta m_d$ value is found to be
$\Delta m_d = 0.509 \pm 0.010\; (\mathrm{stat.}) \pm 0.016\; (\mathrm{syst.})
\;\mathrm{ps}^{-1}$, which agrees well with the world average~\cite{dmd}.

CDF yields a combined opposite-side tagging performance of $\epsilon 
\mathcal D^2 = 1.8\%$, which is an improvement of 20\% with respect 
to the previous CDF analysis~\cite{cdfPRL1}.

Same-side flavor tags are based on the charges of associated 
particles produced in the fragmentation of the $b$ quark that 
produces the reconstructed $B_s$. Contrary to the opposite-side 
tagging algorithms, the performance of this tagging algorithm 
can not be calibrated on $B_d$ and $B^+$ data, but we have to rely 
on Monte Carlo samples until a significant $B_s$ mixing signal has 
been established. CDF uses Neural Network techniques to combine 
kaon particle identification variables from $dE/dx$ measurements 
in the drift-chamber and time-of-flight measurements with 
kinematic quantities of the kaon candidate into a single tagging 
variable. Tracks close in phase space to the $B_s$ candidate are 
considered as same-side kaon tag candidates, and the track with 
the largest value of the tagging variable is selected as the 
tagging track. We predict the dilution of the same-side tag using 
simulated data samples generated with the PYTHIA~\cite{pythia} Monte Carlo 
program. The predicted fractional gain in $\epsilon \mathcal D^2$
from using the Neural Network is 10\%. Control samples of $B^+$ 
and $B_d$ are used to validate the predictions of the simulation. 
The tagging power of this flavor tag is  $\epsilon \mathcal D^2 
= 3.7(4.8)\%$ for the hadronic (semileptonic) decay sample. 
If both a same-side tag and an opposite-side tag are present, 
we combine the information from both tags assuming they are independent.

\subsection{Fit and Results}

An unbinned maximum likelihood fit is used to search for 
$B_s^0-\overline{B}_s^0$  oscillations. The likelihood combines mass, 
proper decay time, proper decay time resolution and flavor 
tagging information for each candidate. Separate probability 
density functions are used to describe signal and each type 
of background. The amplitude scan method~\cite{scan}  was used to 
search for oscillations. The likelihood term describing the 
proper decay time of tagged $B_s$ meson candidates in 
Eq.~\ref{eq:mixProb} is modified by introducing the amplitude $\mathcal A$:
\begin{equation}
  \mathcal L \sim 1\pm \mathcal {AD}\cos(\Delta mt).
\label{eq:like}
\end{equation}
Then, a scan in $\Delta m$ is performed by fitting $\mathcal A$
for fixed values of $\Delta m$. The dilution $\mathcal D$ is  fixed 
to the value obtained by the calibration process. This procedure 
corresponds to a Fourier transformation of the proper time space 
into the frequency space. In the case of infinite statistics and 
perfect resolution, it is expected to find $\mathcal A = 1$ for the 
vicinity of true value of $\Delta m$ and $\mathcal A = 0$ otherwise. 
In practice, the procedure consists in recording $(\mathcal A,\sigma_{\mathcal A})$ 
for each $\Delta m$ hypothesis. A particular value of $\Delta m$ is 
excluded at 95\% C.L. if $\mathcal A + 1.645 \sigma_{\mathcal A} < 1$
holds. The sensitivity of a mixing analysis is defined as the lowest 
$\Delta m$ value for which $1.645 \sigma_{\mathcal A} = 1$.

\begin{figure}[ht]
\centering
\includegraphics[width=80mm]{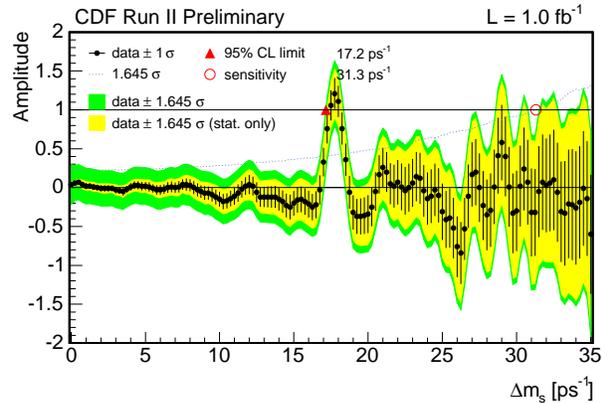}
\caption{Amplitude scan of the hadronic and semileptonic decay modes combined.}
\label{fig:scan}
\end{figure}

The result of the combined amplitude scan for the analysis of the hadronic and 
semileptonic $B_s$ candidates is shown in Fig.~\ref{fig:scan}. 
The combined sensitivity is 
31.3~ps$^{-1}$. The value of the amplitude is consistent with unity 
around $\Delta m_s = 17.75$~ps$^{-1}$, where $\mathcal A = 1.21 \pm 0.20$. 
Elsewhere, the amplitude is always consistent 
with zero (Fig.~\ref{fig:scan}). The minimum likelihood ratio $\Lambda$
is at $\Delta m_s = 17.77$~ps$^{-1}$ and has a value of -17.26. The 
significance of the signal is given by the probability that 
randomly tagged data would produce a value of $\Lambda$ lower than -17.26 
at any value of $\Delta m_s$. Only 28 out of 350 million generated toy 
experiments yielded a $\Lambda$  value lower than that. This results 
in a p-value of $8\times10^{-8}$  which corresponds to a $5.4\sigma$ 
signal. The fit for $\Delta m_s$, with $\mathcal A$ fixed to 
unity, finds
\begin{equation}
  \Delta m_s = 17.77\pm0.10(\mathrm{stat.})\pm0.07(\mathrm{syst.})\;\mathrm{ps}^{-1}.
\label{eq:dms}
\end{equation}
The dominant contributions to the systematic uncertainties comes from 
uncertainties on the absolute scale of the decay-time measurement.

Combining the measured $\Delta m_s$ value with the well known $\Delta m_d$ 
value CDF derive the following ratio of the CKM matrix elements:
\begin{equation}
  \left|\frac{V_{td}}{V_{ts}}\right|=0.2060 \pm 
  0.0007(\mathrm{exp.})^{+0.0080}_{-0.0060}(\mathrm{theor.}).
\label{eq:ckm}
\end{equation}

\section{Observation of New \boldmath{$\Sigma_b$} Baryon}

Until recently only one bottom baryon, the $\Lambda_b^0$, has been 
directly observed. At present the CDF collaboration has accumulated the 
world's largest data sample of bottom baryons, due to a combination 
of two factors: the CDF displaced track trigger, and the $\sim1$ fb$^{-1}$ 
of integrated luminosity delivered by the Tevatron. Using a sample 
of fully reconstructed $\Lambda_b^0\to\Lambda_c^+\pi^-$ candidates 
collected with the displaced track trigger, CDF searched for the decay
$\Sigma_b^{(*)\pm}\to\Lambda_b^0\pi^\pm$.  

\begin{figure}[ht]
\centering
\includegraphics[width=80mm]{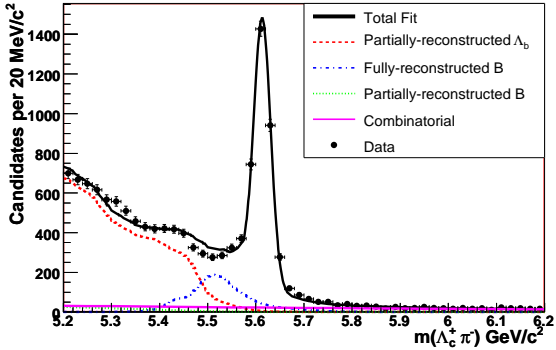}
\caption{Fit to the invariant mass of $\Lambda_b^0\to\Lambda_c^+\pi^-$ 
candidates. The discrepancies between the fit and data below the $\Lambda_b^0$
signal region are due to incomplete knowledge of the branching ratios of the 
decays in this region and are included in the $\Sigma_b^{(*)}$ background 
model systematics.}
\label{fig:Lb}
\end{figure}

CDF reconstructs the decay chain $\Lambda_b^0\to\Lambda_c^+\pi^-\;, 
\Lambda_c^+\to pK^-\pi^+$, reaching a $\Lambda_b^0$ yield of 
approximately 2800 candidates in the signal region $m(\Lambda_b^0) \in
[5.565, 5.670]$~GeV/c$^2$. The $\Lambda_b^0$ mass plot is shown in 
Fig.~\ref{fig:Lb}.

To separate out the resolution on the mass of each $\Lambda_b^0$ candidate, 
CDF searches for narrow resonances in the mass difference 
distribution of $Q = m(\Lambda_b^0\pi) -  m(\Lambda_b^0) -m_\pi$. 
Unless explicitly stated, $\Sigma_b$ refers to both the 
$J = \frac12(\Sigma_b^\pm)$ and $J = \frac32(\Sigma_b^{*\pm})$ states. 
There is no transverse momentum cut applied to the pion from the  
$\Sigma_b$ decay, since these tracks are expected to be very soft. In 
order to perform an unbiased search, the cuts for the $\Sigma_b$
reconstruction are optimized first with the $\Sigma_b$ signal region 
blinded. From theoretical predictions the $\Sigma_b$ signal region is 
chosen as $30 < Q < 100$~MeV/c$^2$, while the upper and lower sideband 
regions of $0 < Q < 30$~MeV/c$^2$ and $100 < Q < 500$~MeV/c$^2$ represent 
the $\Sigma_b$ background. The signal for the optimization is taken from 
a PYTHIA Monte Carlo $\Sigma_b$ sample, with the decays $\Sigma_b \to 
\Lambda_b^0\pi,\; \Lambda_b^0\to\Lambda_c^+\pi^-,\; \Lambda_c^+\to pK^-\pi^+$
forced.

The backgrounds under the $\Lambda_b^0$ signal region in the $\Lambda_b^0$
mass distribution will also be present in the $\Sigma_b$ $Q$-distribution. 
The primary sources of background are $\Lambda_b^0$  hadronization and 
underlying event, hadronization and underlying event of other $B$ meson 
reflections and combinatorial background underneath the $\Lambda_b^0$ peak. 
The percentage of each background component in the $\Lambda_b^0$  signal 
region is derived from the $\Lambda_b^0$ mass fit, and is determined as 
86\% $\Lambda_b^0$  signal, 9\% backgrounds and 5\% combinatorial background.
Other backgrounds (e.g. from 5-track decays where one track is taken as 
the $\pi_{\Sigma_b}$ candidate) are negligible, as confirmed in inclusive 
single-$b$-hadron Monte Carlo samples.

\begin{figure}[ht]
\centering
\includegraphics[width=80mm]{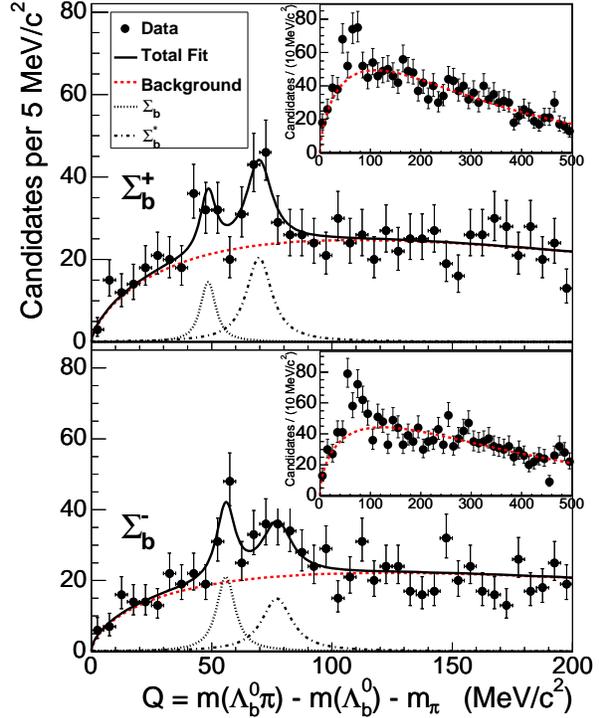}
\caption{Simultaneous fit to the $\Sigma_b$ states. Top plots cotains 
$\Sigma_b^{(*)+}$ states, while the bottom plots contains $\Sigma_b^{(*)-}$
states. The insets show the expected background plotted on the data, while
the signal fit is shown on a reduced range of $Q$.}
\label{fig:sigmab}
\end{figure}

Upon unblinding the $Q$ signal region, there is an excess observed in 
data over predicted backgrounds. CDF performs a simultaneous unbinned 
likelihood fit to ``same charge'' and ``opposite charge'' data. To the 
already described background components, four peaks are added, one for 
each of the expected $\Sigma_b$ states. Each peak is described by a 
non-relativistic Breit-Wigner convoluted with two Gaussian resolution 
functions. The detector resolution has a dominant narrow core and a small 
broader shape describing the tails where the PDF for each peak takes both 
into account. Due to low statistics, CDF constrains $m(\Sigma_b^{*+}) - 
m(\Sigma_b^+)$ and $m(\Sigma_b^{*-}) - m(\Sigma_b^-)$ to be the same. 
The results of the fit are displayed in Fig.~\ref{fig:sigmab}.

All systematic uncertainties on the mass difference measurements are 
small compared to their statistical errors.

To summarize, the lowest lying charged $\Lambda_b^0\pi$ resonant states 
are observed in 1~fb$^{-1}$ of data collected by the CDF detector. 
These are consistent with the lowest lying charged $\Sigma_b^{(*)\pm}$
baryons. 
Using the best CDF mass measurement for the $\Lambda_b^0$ mass, which is 
$m(\Lambda_b^0) = 5619.7\pm1.2(\mathrm{stat.})\pm1.2(\mathrm{syst.})$~MeV/c$^2$, 
the absolute mass values are measured to be:
\begin{eqnarray}
  m(\Sigma_b^-)    & = & 5815.2\pm1.0(\mathrm{stat.}) \pm 
  1.7(\mathrm{syst.})\;\mathrm{MeV}  \nonumber\\
  m(\Sigma_b^+)    & = & 5807.8^{+2.0}_{-2.2}(\mathrm{stat.}) \pm 
  1.7(\mathrm{syst.})\;\mathrm{MeV}  \nonumber\\
  m(\Sigma_b^{*-}) & = & 5836.4\pm2.0(\mathrm{stat.})  
  ^{+1.8}_{-1.7}(\mathrm{syst.})\;\mathrm{MeV}  \nonumber\\
  m(\Sigma_b^{*+}) & = & 5829.0^{+1.6}_{-1.8}(\mathrm{stat.})  
  ^{+1.7}_{-1.7}(\mathrm{syst.})\;\mathrm{MeV}  \nonumber
\label{eq:SbMass}
\end{eqnarray}

\section{Lifetimes Measurements in \boldmath{$J/\psi$} Decays}

\begin{figure*}[hbt]
\centering
\includegraphics[width=78mm]{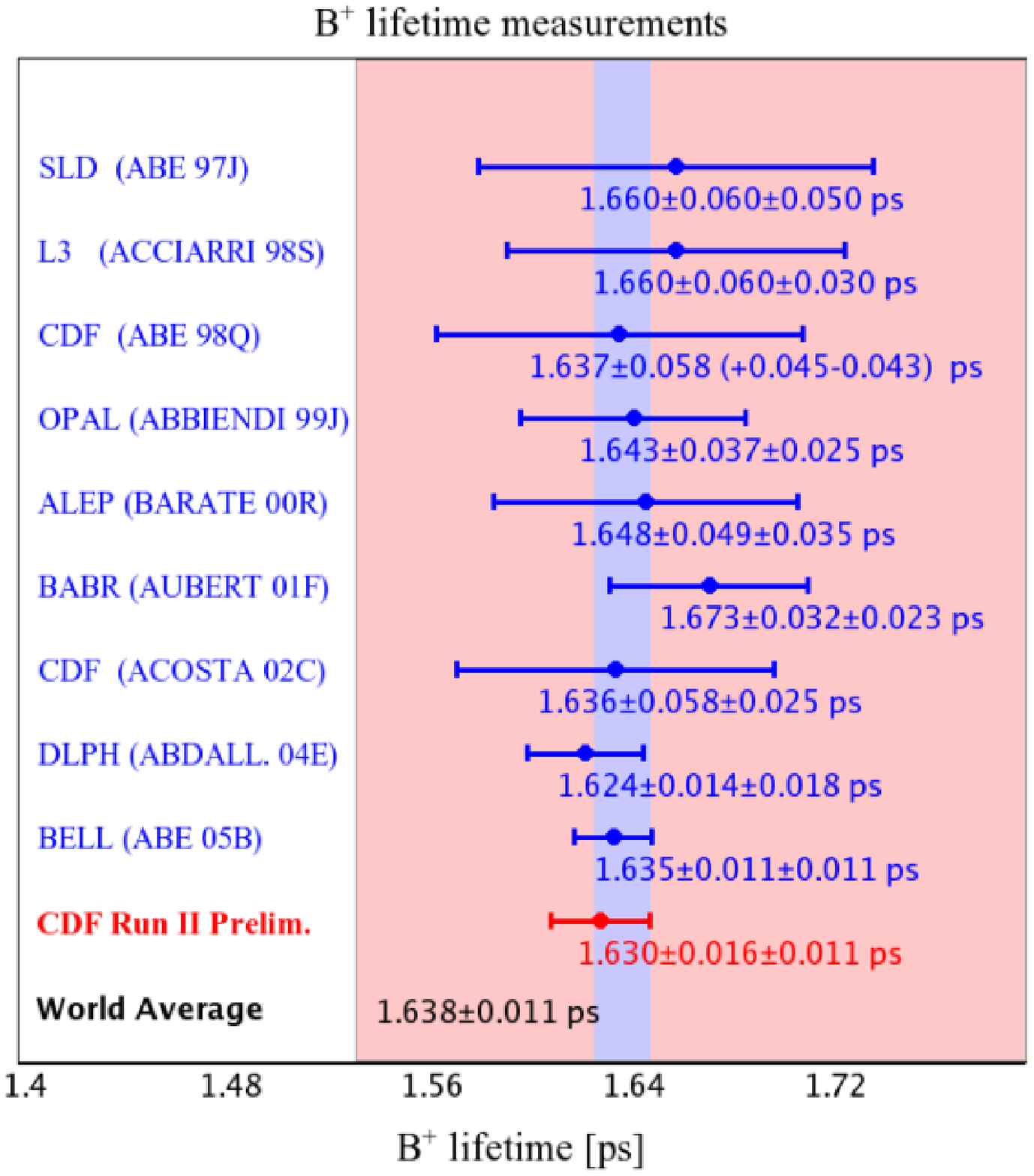}\hspace{10mm}
\includegraphics[width=78mm]{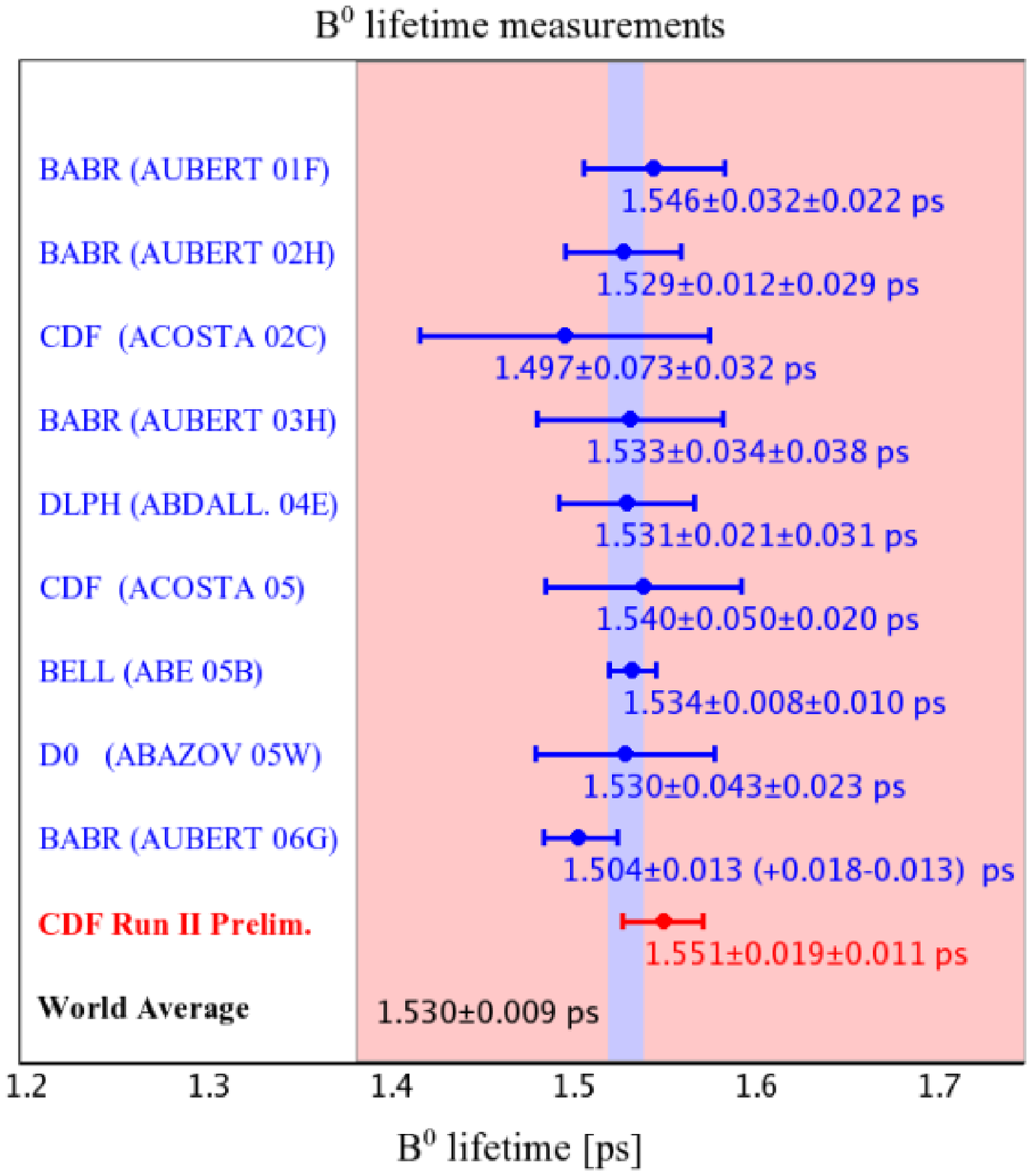}\vspace{5mm}
\includegraphics[width=78mm]{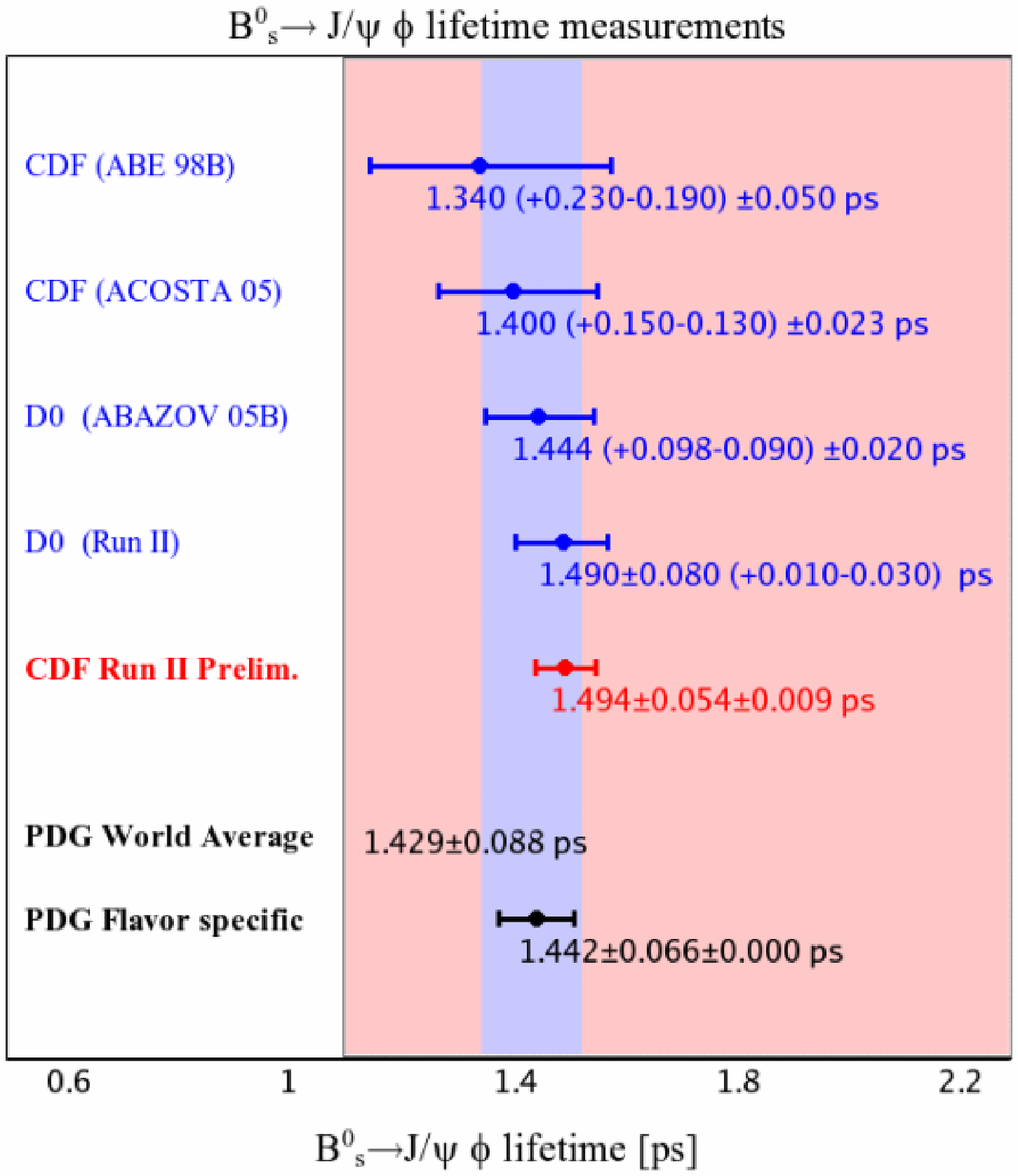}\hspace{10mm}
\includegraphics[width=78mm]{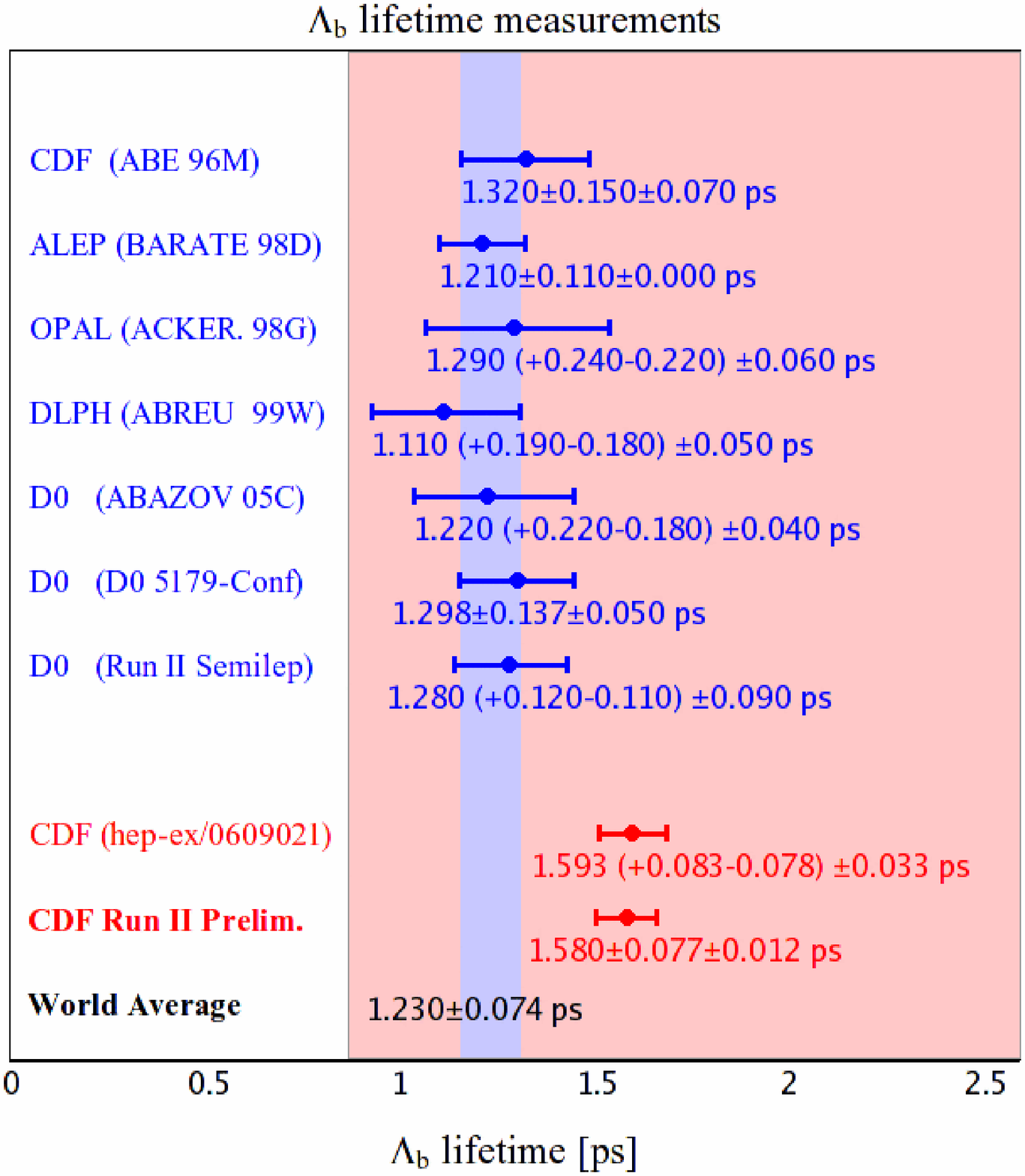}
\caption{Comparison of measured lifetimes with a selection of results
quoted in the PDG2006 and others. Note: the world average values are from
PDG2006 and do not include the CDF preliminary results.}
\label{fig:lifetimes}
\end{figure*}

In a simple quark spectator model, the lifetime of a $B$ hadron
is governed by the decay of the $b$-quark, and the lifetimes of 
all $B$ hadrons are expected to be the same. However, because of 
significant non-spectator effects, the $B$ hadron lifetimes 
follow a hierarchy: $\tau(B^+) \ge \tau(B^0) \sim \tau(B^0_s) > 
\tau(\Lambda_b^0) \gg \tau(B^+_c)$. 
This hierarchy is predicted by the Heavy Quark Expansion (HQE) 
technique~\cite{hqe}, which expresses decay widths of heavy hadrons 
as an expansion in inverse powers of the heavy quark mass (i.e. 1/$m_b$). 

CDF presents an updated measurement of exclusive $B$ lifetimes in
the modes $B^+\to J/\psi K^+$, $B^0\to J/\psi K^{*0}$, $B^0\to J/\psi K_s^0$,
$B_s\to J/\psi\phi$ and $\Lambda_b\to J/\psi \Lambda$, based on 
1.0~fb$^{-1}$ of integrated luminosity collected with the di-muon trigger.

Signal yields for all decay channels and measured lifetimes for $B^+$, $B^0$,
$B_s$ and $\Lambda_b$ are summarized in Table~\ref{tab:lifetimes} and 
compared with other experiments and PDG2006 average~\cite{PDG2006} in 
Fig.~\ref{fig:lifetimes}. 

\begin{table}[h]
\begin{center}
\caption{Signal yields for all the channels and measured lifetimes. First 
uncertainty is statistical and the second is systematic.}
\begin{tabular}{|l|c|c|}
  \hline
  Decay Channel                  & Signal Yield & Lifetime [ps]         \\ \hline
  $B^+\to J/\psi K^+$            &     12,900   &                 $1.630\pm0.016\pm0.011$ \\ \hline
  $B^0\to J/\psi K^{*0}$         &     4,800    & \multirow{2}{*}{$1.551\pm0.019\pm0.011$}\\ \cline{1-2}
  $B^0\to J/\psi K_s^0$          &     3,600    &                                         \\ \hline
  $B_s\to J/\psi\phi$            &     1,100    &                 $1.494\pm0.054\pm0.009$ \\ \hline
  $\Lambda_b\to J/\psi \Lambda$  &     530      &                 $1.580\pm0.077\pm0.012$ \\ \hline
\end{tabular}
\label{tab:lifetimes}
\end{center}
\end{table}

Results of the $B^+$ and $B^0$ mesons lifetimes are in good agreement with the 
world average, with uncertainties that are comparable to individual 
uncertainties from $B$ factories results. The measured lifetime for
the $B_s$ meson also agrees well with the world average, and its uncertainty
is more precise than the global uncertainty from the world average. 

The $\Lambda_b$ lifetime result is the most precise measurement to date.
It is consistent with most of individual results from other experiments, although 
it is $\sim3\sigma$ above the world average. An independent recent result from 
CDF has also shown a similar trend above the world average.

\section{\boldmath{$B\to\mu^+\mu^-h$} Searches}

The decay of a $b$ quark into an $s$ quark and two muons requires a 
flavor-changing neutral current (FCNC) process which is strongly suppressed
in the standard model. New physics models allow for significant deviations
from the standard model prediction. While the $b\to s\gamma$ branching ratio 
has been accurately measured~\cite{PDG2006} and agrees with the theory predictions,
the $b\to s\mu^+\mu^-$ transition allows the study of FCNC in more detail
through additional observables, such as the dimuon invariant mass, and the
forward-backward asymmetry of the strange quark in the dimuon system.

The rare decays $B^+\to\mu^+\mu^-K^+$ and $B^0\to\mu^+\mu^-K^{*0}$ have
been observed at the $B$ factories~\cite{babar,belle}. However, searches for 
the analogous $B_s\to\mu^+\mu^-\phi$ decay, with a predicted branching ratio
of $1.6\times10^{-6}$~\cite{br}, have not revealed a significant signal.

CDF search in 924~pb$^{-1}$ of data for the rare decay modes $B^+\to\mu^+\mu^-K^+$,
$B^0\to\mu^+\mu^-K^{*0}$ and $B_s\to\mu^+\mu^-\phi$. The $K^*$ is reconstructed 
in the mode $K^*\to K^+\pi^-$, and the $\phi$ is reconstructed as $\phi\to K^+K^-$.

The offline analysis begins by searching for a pair of oppositely charged 
muon tracks. The two muon tracks are combined with a third charged track 
from a $B^+\to\mu^+\mu^-K^+$ candidate, or another pair of oppositely 
charged tracks from a $B^0\to\mu^+\mu^-K^{*0}$ or $B_s\to\mu^+\mu^-\phi$ \
candidate. We exclude events where the dimuon invariant mass is within 
the $J/\psi\to\mu^+\mu^-$ and $\psi(2S)\to\mu^+\mu^-$ mass regions to 
eliminate possible contributions from the charmonium resonant decays. 

Muons are required to have $p_T > 1.5$ or $2.0$~GeV/c depending on which dimuon 
trigger selected the event. The kaon requirement is $p_T > 0.4$~GeV/c. 
The following three discriminating variables are used in the optimization 
of the searches: the proper lifetime significance, $ct/\sigma_{ct}$; the 
pointing angle $\alpha$ from the $B$ meson candidate to the primary vertex; and
the isolation, $I$, defined as the transverse momentum carried by the $B$ meson
candidate divided by the transverse momentum of all charged particles in a 
cone  around the direction of the $B$ meson candidate.
The expected number of background events in the $B$ mass window is 
obtained by extrapolating events in the high-mass sideband to 
the signal region. Since the region below the $B$ signal window 
contains partially reconstructed $B$ decays, only the high-mass 
sideband is used in the background estimate. The figure-of-merit 
for the optimization is $S/\sqrt{(S + B)}$, where $S$ is the estimate 
of the expected yield of the rare decays, and $B$ is the expected 
background. For the $B^+$ and $B^0$ rare decay searches, the PDG 
values of the branching fractions are used in the optimization, 
while the theoretical expectation is used for the $B_s$ search. 
The optimization is performed separately for the three rare 
decay modes. The resulting optimal values are very similar 
between the different modes and the following averages are used 
for all three searches: $ct/\sigma_{ct} > 14$, $\alpha < 0.06$, and 
$I > 0.6$. 

\begin{figure*}[ht]
\centering
\includegraphics[width=55mm]{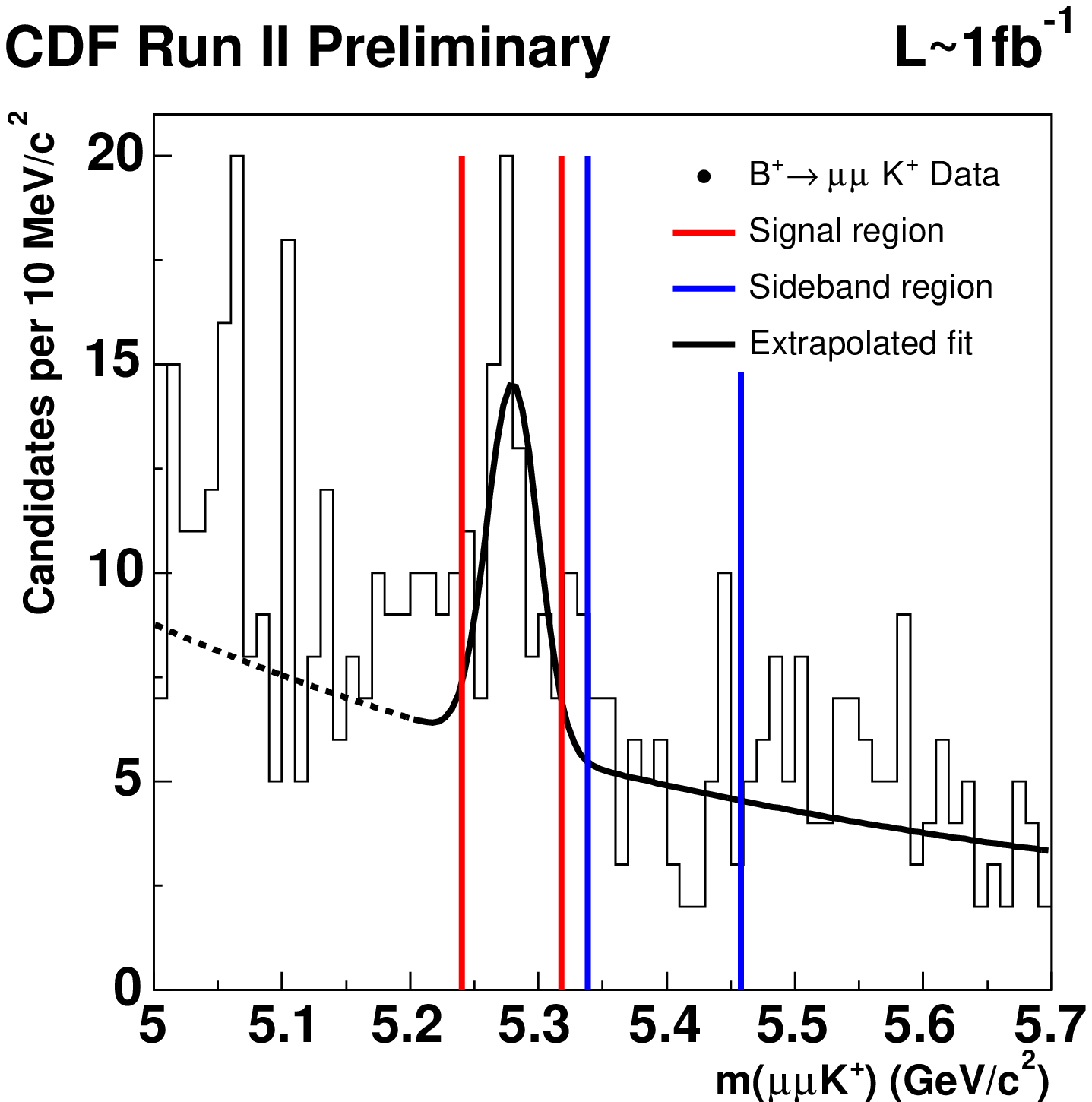}
\includegraphics[width=55mm]{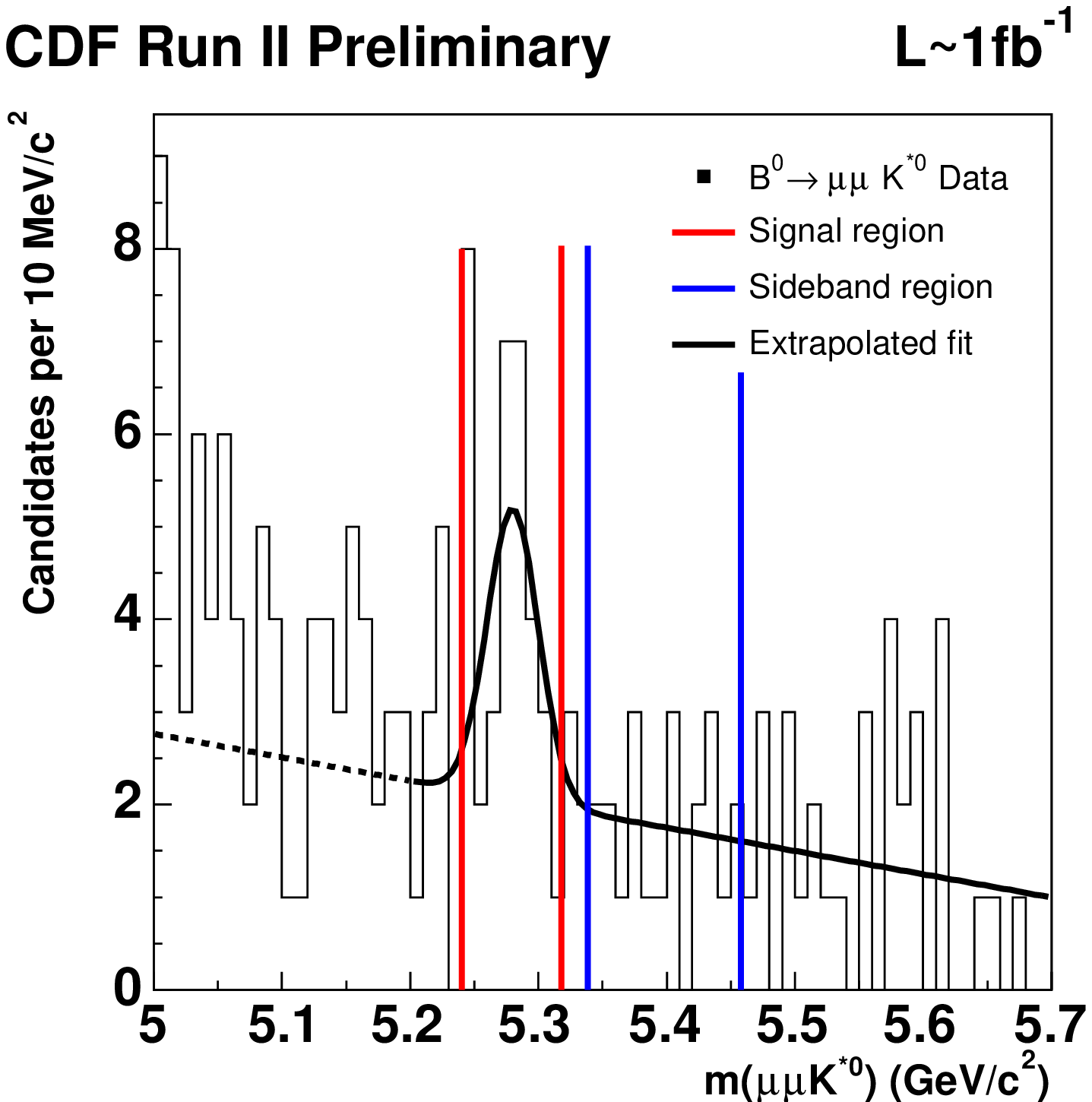}
\includegraphics[width=55mm]{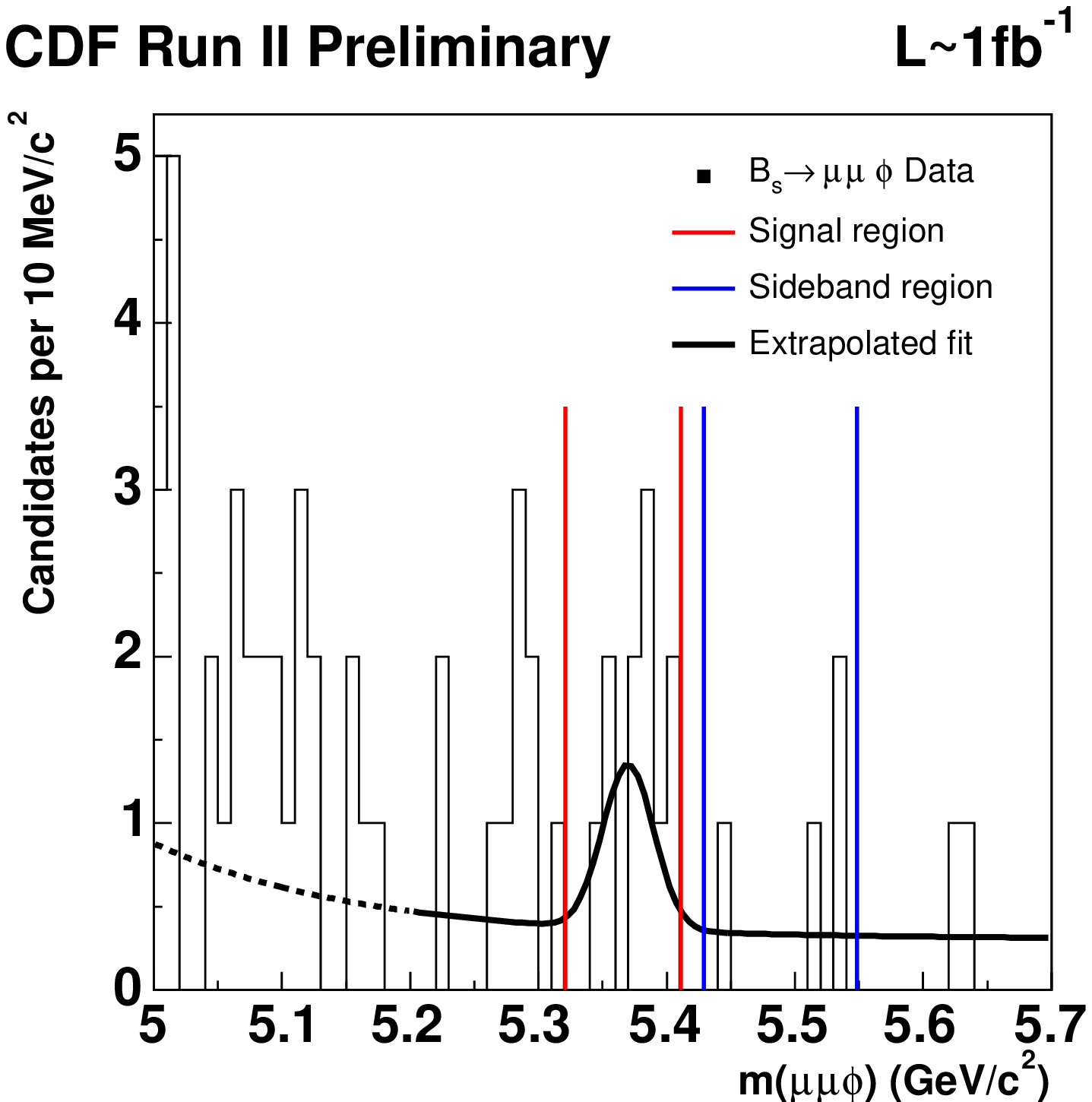}
\caption{The invariant distribution for the three rare decay modes. 
The vertical bars define the signal and sideband regions. The black 
curve illustrates the expected shape for the signal and combinatoric 
background.}
\label{fig:mumu}
\end{figure*}

The invariant mass distribution for the three searches 
after applying the optimal requirements on the discriminating 
variables are shown in Fig.~\ref{fig:mumu}. An excess is found in all three
modes. The significance of each excess is determined by calculating the 
probability for the background to fluctuate into the number of observed
events or more. A significance of 4.5, 2.9, and 2.4 standard deviations
is found respectively for the $B^+$, $B^0$ and $B_s$ modes.

The branching fraction can be computed by normalizing the number of 
the observed signal to the number of reconstructed resonant $B\to J/\psi h$ 
decays:
\begin{equation}
  \frac{\mathcal B(B\to\mu^+\mu^-h)}{\mathcal B(B\to J/\psi h)} =
  \frac{N_{\mu^+\mu^-h}}{N_{J/\psi h}}\frac{\epsilon_{\mu^+\mu^-h}}{\epsilon_{J/\psi h}}
  \times\mathcal B(J/\psi\to\mu^+\mu^-),
\label{eq:norm}
\end{equation}
where $h$ stands for $K^+$, $K^*$, or $\phi$. The parameter $N_{\mu^+\mu^-h}$
is either the number of observed signal events or, in the case of setting a 
limit, the upper limit on the number of signal decays, and $N_{J/\psi h}$ 
is the number of reconstructed $B\to J/\psi h$ events. The efficiency terms 
$\epsilon_{\mu^+\mu^-h}$ and $\epsilon_{J/\psi h}$ are the efficiency for 
reconstructing the normalization and signal decays, respectively.

Using the world average branching ratio of the normalization modes~\cite{PDG2006},
we extract the following branching ratios using Eq.~\ref{eq:norm}:
\begin{eqnarray}
  \mathcal B(B^+\to\mu^+\mu^-K^{+})  & = & (0.60\pm0.15\pm0.04)\times10^{-6},\nonumber\\
  \mathcal B(B^0\to\mu^+\mu^-K^{*0}) & = & (0.82\pm0.31\pm0.10)\times10^{-6}, \nonumber
\end{eqnarray}
first uncertainty is statistical and second systematic. These measurements are
consistent with the world average and of similar precision as the best available 
measurements.

Since the $B_s\to\mu^+\mu^-\phi$ excess is not significant, we calculate a 
limit on its relative branching ratio using a Bayesian approach. We find:
\begin{equation}\nonumber
  \frac{\mathcal B(B^0_s\to\mu^+\mu^-\phi)}{\mathcal B(B^0_s\to J/\psi\phi)} < 
  2.30\times10^{-3}\;\;\mathrm{at \;90\% \; C.L.} 
\end{equation}
This limit on the $B_s$ mode is the most stringent to date.

\section{Charmless Two-Body \boldmath{${B}$} Decays: 
  \boldmath{${B^0\to h^+h^{\,\prime-}}$}}

The decay modes of $B$ mesons into pairs of charmless pseudoscalar mesons are
effective probes of the quark-mixing (CKM) matrix and are sensitive to
potential new physics effects. The large production of $B$ hadrons of all
kinds at the Tevatron allows an opportunity for measuring such decays in new 
modes, which are important to supplement our understanding of $B$ meson
decays.

\subsection{Event Selection and Fit of Composition}

$B$ Hadrons are initially selected by using the two-track trigger. 
In the offline analysis, additional cuts are imposed on isolation, 
$I$ --defined previously--, and the quality of the fit, $\chi^2$,  
to the 3D decay vertex of the $B$ hadron candidate. Final selection 
cuts are determined by an optimization procedure, based on minimizing 
the expected uncertainty of the physics observables to be measured. 
Two different sets of cuts are used, optimized respectively for best 
resolution on $A_{CP}(B^0\to K^+\pi^-)$ (loose cuts), and for best 
sensitivity for the discovery of the yet unobserved $B^0_s\to K^-\pi^+$ 
mode (tight cuts). The looser set of cuts is also used for measuring 
the decay rates of the largest yield modes, while the tighter 
set is used for the other rare modes. 

\begin{figure}[ht]
\centering
\includegraphics[width=80mm]{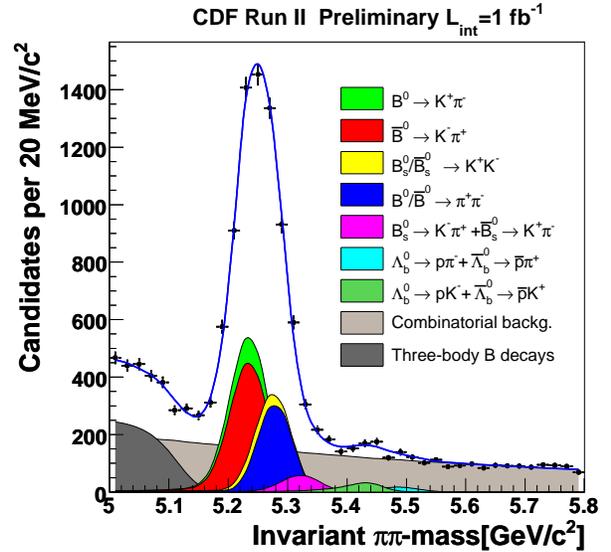}
\caption{Invariant mass distribution of ${B^0\to h^+h^{\,\prime-}}$ candidates
passing the loose selection cuts. The pion mass is assigned to both tracks.}
\label{fig:mpipi}
\end{figure}

The invariant mass distribution of the candidates, with an arbitrary 
pion mass assignment to both tracks, shows a single large peak in 
the $B$ mass range (Fig.~\ref{fig:mpipi}), formed by several 
overlapping modes.

The different modes are statistically separated and individually 
measured by means of an unbinned maximum-Likelihood fit, combining 
kinematics and PID. Kinematic information is summarized by three 
loosely correlated observables: the mass $M_\pi\pi$ calculated 
with the pion mass assignment to both particles; the signed momentum 
imbalance $\alpha = (1 - p_1/p_2)q_1$, where $p_1$ ($p_2$) 
is the lower (higher) of the particle momenta, and $q_1$ is the 
sign of the charge of the particle of momentum $p_1$; the 
scalar sum of particle momenta $p_{tot} = p_1 + p_2$. The above 
variables allow evaluating the mass of the $B$ candidate 
for any mass assignment to the decay products. PID information 
is given by a $dE/dx$ measurement for each track.

The shape of the mass distribution of each single channel accounts 
for non-Gaussian tails, both from resolution and from emission of 
photons in the final state, which is simulated on the basis of 
analytical QED calculations~\cite{qed}. The mass distribution
of the combinatorial background is fit to a smooth function, while 
the physics background is parameterized by an 'Argus function'~\cite{argus} 
smeared with our mass resolution. Kinematical distributions for 
the signal are represented by analytical expressions, while for 
the combinatorial background are parameterized from the mass sidebands 
of data.

The dominant contributions to the systematic uncertainty come from: 
statistical uncertainty on isolation efficiency ratio (for $B^0_s$ modes); 
uncertainty on the $dE/dx$ calibration and parameterization; and uncertainty 
on the combinatorial background model. Smaller systematic uncertainties 
are assigned for: trigger efficiencies; physics background shape and 
kinematics; $B$ meson masses and lifetimes.

\subsection{Results}

The search for rare modes is performed using the tight selection. 
The fit allows for the presence of any component of the form 
$B\to h^+h^{\,\prime-}$ or $\Lambda_b^0\to ph^-$ where $h$, $h' = K$ 
or $\pi$, with the yield as a free parameter.

The results provide the first observation of the $B^0_s\to K^-\pi^+$ mode, 
with a significance of $8.2\sigma$, which includes systematic uncertainties 
and is evaluated from Monte Carlo samples of background without signal. 
The branching fraction of this mode is significantly sensitive to 
the value of angle $\gamma$ of the unitary triangle. Our measurement 
$\mathcal B(B^0_s\to K^-\pi^+) = (5.0 \pm 0.75 \pm 1.0) \times10^{-6}$ 
is in agreement with the prediction in~\cite{williamson} , but is lower 
than most other predictions~\cite{beneke,yu,sun}.

No evidence is found for modes $B_s^0\to\pi^+\pi^-$ or $B^0\to K^+K^-$, 
in agreement with expectations of significantly smaller branching fractions.
An upper limit for the branching ratio on these decay modes is set:
\begin{eqnarray}
  \mathcal B(B^0\to K^+K^-)     & < 0.7\times 10^{-6} & \mathrm{\;at\; 90\%\; CL}, \nonumber\\
  \mathcal B(B_s^0\to \pi^+\pi^-) & < 1.36\times10^{-6} & \mathrm{\;at\; 90\%\; CL}. \nonumber
\end{eqnarray}

In the same sample, we also get to observe charmless decays of a $B$ 
baryon for the first time: $\Lambda_b^0\to p\pi^-$ ($6\sigma$) and 
$\Lambda_b^0\to pK^-$ ($11.5\sigma$). We measure the ratio of branching 
fractions of these modes as $\mathcal B(\Lambda_b^0\to p\pi^-)/
\mathcal B(\Lambda_b^0\to pK^-) = 0.66 \pm 0.14 \pm 0.08$, in good agreement 
with the expected range [0.60, 0.62] from~\cite{lambdaBR}.

We can measure from our data the asymmetries of both $B^0$ and $B_s^0$ 
decays in the self-tagging final state $K^\pm\pi^\mp$ . The asymmetry of the 
$B_s^0$ mode is measured with the tight selection, while the looser selection 
is used for the $B^0$ mode. 

The result $A_{CP}(B^0\to K^+\pi^-) = -0.086\pm0.023\pm0.009$ is in agreement 
with the world average~\cite{hfag}, and is the second most precise measurement. 

Using the tight set of cuts CDF is able to achieve the first CP asymmetry 
measurement on the $B_s^0\to K^-\pi^+$ system, finding 
$A_{CP}(B_s^0\to K^-\pi^+) = 0.39 \pm 0.15 \pm 0.08$. 
This value favors the large CP asymmetry predicted by the Standard Model 
and has the correct sign~\cite{gronau}, but is still compatible with zero (significance 
just above $2\sigma$).

\section{Conclusions}

The heavy flavor physics program at CDF is being very productive. We have reviewed 
some of the most recent CDF results which make use of $\sim1$~fb$^{-1}$.
These results include the observation of the $B_s$ oscillation frequency, the first 
observation of bottom baryon $\Sigma_b^{(*)\pm}$ states, updates on $B$ 
hadrons lifetimes, and searches for rare decays in the $b\to s\mu^+\mu^-$ 
transition and in charmless two-body $B$ decays. CDF achieve complementary 
and competitive results with $B$ Factories, being some of them unique at CDF.
With 2.5~fb$^{-1}$ already on tape we expect more and new interesting results
for this summer.

\begin{acknowledgments}

The results shown here represent the work of many people.
I would like to thank all colleagues from CDF for their efforts to carry 
out these challenging physics analyses, the conference organizers for 
a very nice days of physics, and the colleagues of my research
institution, IFCA, for all their help. 

\end{acknowledgments}

\bigskip 

\end{document}